\documentclass[12pt,a4paper,english]{article}
\usepackage[T1]{fontenc}
\usepackage[latin2]{inputenc}
\usepackage[english]{babel}
\usepackage{amsmath}
\usepackage{amsfonts}
\usepackage{indentfirst}
\usepackage[dvips]{graphicx}
\usepackage{youngtab}

\newcommand{\bea}{\begin{eqnarray}}
\newcommand{\eea}{\end{eqnarray}}
\newcommand{\be}{\begin{equation}}
\newcommand{\ee}{\end{equation}}

\newcommand{\hf}{\frac{1}{2}}

\newcommand{\Z}{{\mathbb Z}}
\newcommand{\R}{{\mathbb R}}
\newcommand{\C}{{\mathbb C}}

\def\Tr{{\rm Tr}}

\def\G{\Gamma}

\newcommand{\cF}{{\cal F }}

\begin{document}

\sloppy


\begin{flushright}
\begin{tabular}{l}
ITFA-2007-26\\
IFT-UW-2007-5 \\
BONN-TH-2007-12\\

\\ [.3in]
\end{tabular}
\end{flushright}

\begin{center}
\Large{ \bf Instantons on ALE spaces and orbifold partitions}
\end{center}

\begin{center}

\bigskip

Robbert Dijkgraaf $^{1,2}$ and Piotr Su\l kowski $^{1,3,4,5}$

\bigskip

\emph{$^1$Institute for Theoretical Physics and $^2$KdV Institute for Mathematics} \\    
\emph{University of Amsterdam, Valckenierstraat 65, 1018 XE Amsterdam, The Netherlands } \\[4mm]
\emph{$^3$\!\! Institute for Theoretical Physics, Warsaw University,} \\
\emph{and \,$^4$\!\! So{\l}tan Institute for Nuclear Studies,} \\
\emph{ul. Ho\.za 69, 00-681 Warsaw, Poland} \\ [4mm]
\emph{$^5$Physikalisches Institut der Universit{\"a}t Bonn,} \\
\emph{Nussallee 12, 53115 Bonn, Germany} \\

\bigskip

\centerline{ \emph{R.H.Dijkgraaf@uva.nl, Piotr.Sulkowski@fuw.edu.pl} }

\smallskip
 \vskip .6in \centerline{\bf Abstract}
\smallskip

\end{center}

We consider $\mathcal{N}=4$ theories on ALE spaces of $A_{k-1}$ type. As is
well known, their partition functions coincide with $A_{k-1}$ affine
characters. We show that these partition functions are equal to the generating
functions of some peculiar classes of partitions which we introduce under the
name 'orbifold partitions'. These orbifold partitions turn out to be related
to the generalized Frobenius partitions introduced by G. E. Andrews some years
ago. We relate the orbifold partitions to the blended partitions 
and interpret explicitly in terms of a free fermion system.


\newpage


\section{Introduction}

There is a rich class of relations between four-dimensional supersymmetric gauge theories, particularly the topological twisted ones that are given by exact instanton sums, and two-dimensional
conformal field theories, which can often be realized in
terms of free fermion systems. 

One of the first examples of such a relation was found by Nakajima
\cite{naka1} in the mathematical literature and was further explained
by Vafa and Witten as a manifestation of S-duality of $\mathcal{N}=4$
theories \cite{V-W}. Various gauge theory quantities such as
partition functions, can be computed by closely
related ensembles of partitions in the corresponding 2d CFT. In
particular, in the simplest case, the partition function of
$\mathcal{N}=4$ twisted $U(1)$ gauge theory on $\mathbb{R}^4$ is given
by the inverse of the Dedekind $\eta(q)$ function, which apart from a
factor of $q^{-1/24}$ is a generating function of all two-dimensional
partitions weighted by the number of boxes.
For $\mathcal{N}=2$ theories on $\R^4$ the free fermion expressions
encoding their partition functions were found and related to the
counting of various ensembles of two-dimensional partitions in
\cite{Nek,Nek-Ok}. In general, these relations arise from an
identification of the gauge theory partition functions with the Euler
characteristics of instanton moduli spaces, with two-dimensional
partitions encoding information about these moduli spaces.

Recently, the appearance of conformal field theories and free fermions
was understood more directly from the (physical type II) string theory
perspective. In string theory supersymmetric gauge theories can be
realized on the worldvolume of D4-branes. It was shown in \cite{dhsv}
(see also \cite{Ibrane}) that such D4-branes wrapping Taub-NUT spaces
can be related to a system of D4 and D6-branes intersecting along a
Riemann surface. Then the fermions in question arise as massless
states of open strings stretched between D4 and D6-branes and indeed
live just on this surface. Moreover, by a suitable chain of dualities
those D-brane systems were related to topological strings.

In fact, the above relations are not limited to four-dimensional gauge
theories. As shown in \cite{foam}, the partition function of
six-dimensional twisted $U(1)$ gauge theory on $\R^6$ is computed by
three-dimensional partitions and is closely related to the topological
vertex formulation of topological strings.

Even though free fermion expressions have been found for gauge
theories defined on various underlying manifolds, their interpretation
in terms of partitions was understood so far mainly in the case of
theories defined on $\R^4$ or $\R^6$. In this paper we extend the
partition interpretation to $\mathcal{N}=4$ theories on ALE spaces,
whose partition functions are well-known to be given by affine
characters \cite{naka1,V-W}. At least for the $U(1)$ theory on an
$A_{k-1}$ singularities we show that their partition functions are
generating functions of special classes of partitions, which we call
\emph{orbifold} or \emph{generalized partitions}. These partitions are
related to the so-called \emph{generalized Frobenius partitions}
introduced by G.E.~Andrews in \cite{Andrews}. We also interpret these
orbifold partitions explicitly in terms of a two-dimensional fermionic
system. The issue of instantons on ALE spaces as well as their relation
to counting of partitions were also analyzed, among the others, in 
\cite{bianchi,ALEspaces,fujii-minabe}.

The paper is organized as follows. In section \ref{ssec-N4} we review
the most essential properties of $\mathcal{N}=4$ twisted Yang-Mills
theory, in particular when defined on ALE spaces. In section
\ref{sec-crystals} we introduce two types of orbifold partitions for
ALE spaces of $A_{k-1}$ type. In sections \ref{sec-crystals-resum-0}
and \ref{sec-crystals-resum} we show how to compute generating
functions of these orbifold partitions and prove they are equal to
appropriate affine characters. In section \ref{sec-examples} we give
some explicit examples of these generating functions. A summary of
results is given in section \ref{summarize}.



\section{$\mathcal{N}=4$ Yang-Mills theory and ALE spaces}  \label{ssec-N4}

In this section we briefly review some properties of $\mathcal{N}=4$
topological supersymmetric $U(N)$ gauge theory following
\cite{V-W,dhsv}. An ordinary $\mathcal{N}=4$ theory can be though of
as $\mathcal{N}=2$ theory with a chiral multiplet and a massless
hypermultiplet. We consider its twisted version constructed by the
embedding of the $SO(4)$ Euclidean rotation group specified by the
$({\bf 1},{\bf 2})\oplus({\bf 1},{\bf 2})$ representation into the
$SU(4)$ $R$-symmetry group. After the twisting the theory becomes
topological, albeit equivalent to an ordinary theory when defined on a
hyper-K{\"a}hler four-manifold $M$, and its action is given by
$$ 
- \int {i\over 8 \pi} \tau\, \Tr\,F_+\wedge F_+ + v \wedge \Tr\,F_+
+ \hbox{\it c.c.}
$$ with $v \in H^2(M,\Z)$, $F_+$ denoting the self-dual part of the
field strength, and the complexified gauge coupling $\tau$ given by
$$
\label{tau}
\tau = {\theta\over 2\pi} + {4\pi i \over g^2}.
$$
The partition function of this theory takes the form
\be
Z(v,\tau) = q^{-N\chi(M)/24} \sum_{m,n} d(m,n) y^m q^n,         \label{Z-gauge}
\ee
where $d(m,n)$ compute the Euler number of instanton moduli space of
topological charges $c_1=m$ and $ch_2=n$, while $y=e^{2\pi i v}$ and
$q=e^{2\pi i \tau}$. It is important to stress that the coefficients
$d(m,n)$ are expected to be integers. This becomes particularly clear
from a five-dimensional perspective, where $d(m,n)$ acquires an
interpretation of an index which computes certain BPS invariants.

The beta-function of $\mathcal{N}=4$ theory vanishes and in
consequence this theory has a very peculiar properties under the
action of the $S$-duality group on the gauge coupling $\tau$. In
particular, the partition function should transform as a modular
Jacobi-form
$$ 
Z\left({v \over c\tau + d},{a\tau +b \over c\tau +d} \right) 
 =  (c\tau + d)^{-\chi(M)} e^{2\pi i c v^2 N/(c\tau +d)}  Z(v,\tau), 
$$
$$
Z(v + n\tau + m, \tau) = e^{-2\pi i N (n^2\tau + 2n \cdot v)} Z(v,\tau), 
$$
$$
\textrm{for}\quad \left[\begin{array}{cc}
a & b \\
c & d \\
\end{array}\right] \in SL(2,\Z), \quad \textrm{and} \quad 
n,m \in H^2(M,\Z) \cong \Z^{b_2}. \nonumber
$$
Such a behavior is a reminiscent of a transformation property of
affine characters. If $Z(v,\tau)$ indeed coincides with some affine
character, this automatically ensures integrality of $d(m,n)$, as in
that case they encode multiplicities of certain representations
specified by affine integrable weights.

\begin{figure}[htb]
\begin{center}
\includegraphics[width=0.4\textwidth]{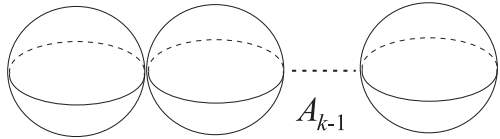}
\begin{quote}\caption{An ALE space of $A_{k-1}$ type contains $k-1$ spheres whose
  intersection numbers are given by the Cartan matrix of the $A_{k-1}$ Lie
  algebra. If all spheres are blow down to zero size, we obtain the singular
  orbifold space $\C^2/\Z_{k}$.} 
\label{fig-ale}
\end{quote}
\end{center}
\end{figure}

In this paper we wish to consider $\mathcal{N}=4$ $U(N)$ theories on
ALE spaces. In general, ALE spaces can be constructed as resolutions
of orbifolds of the form
\be
\mathbb{C}^2/\Gamma,         \label{ale-space}
\ee
where $\Gamma$ is a finite subgroup of $SU(2)$. In the process of a
resolution the singular orbifold points are replaced by two-spheres,
whose intersection numbers are equal to entries of a Cartan matrix of
a certain Lie algebra $\mathfrak g$. This provides a mapping between
two-spheres in the resolved singularity and nodes in a Dynkin diagram
of $\mathfrak g$, as well as a one-to-one correspondence between
finite subgroups $\Gamma$ of $SU(2)$ and Lie algebras $\mathfrak g$
known as the McKay correspondence \cite{mckay}. In particular, under
this correspondence abelian groups $\Gamma=\Z_{k}$ are mapped to
$A_{k-1}$ Lie algebras, dihedral groups are mapped to Lie groups of
$D$ type, and symmetry groups of regular solids are mapped to
exceptional Lie algebras of $E$ type. The ALE space of $A_{k-1}$ type is
shown schematically in figure \ref{fig-ale}.

Aspects of $\mathcal{N}=4$ $U(N)$ theories on ALE spaces were
elucidated in \cite{naka1,V-W,dhsv}. The boundary of the ALE space is
a Lens space $S^3/\Gamma$, so the $U(N)$ gauge field can approach some
non-trivial flat connection at infinity. Such flat connection is
labelled by the $N$-dimensional representation of $\Gamma$
$$
\lambda \in \textrm{Hom} (\Gamma,U(N)),
$$
which can be decomposed into irreducible representations $\rho_i$ of $\Gamma$ as
$$
\lambda = \sum_i N_i \rho_i,
$$
with $N_i$ non-negative integers satisfying the relation
$$
\sum_i N_i\, \textrm{dim}\, \rho_i = N.
$$ 
Under the McKay correspondence the dimensions $d_i$ of the irreducible
representations of $\G$ can be identified with the dual Dynkin indices
of the extended Dynkin diagram. This suggests we can think of each
$\lambda$ as determining an integrable highest-weight representation
of the the affine extension $\widehat{\mathfrak g}$ at level $N$. We
will denote this representation as $V_{\widehat{\lambda}}$, where
$\widehat{\lambda}$ is the corresponding integrable weight. In a
remarkable work \cite{naka1} Nakajima proved that on the middle
dimensional cohomology of the moduli space of gauge theory one can
actually realize the affine algebra $\widehat{\mathfrak g}$, and in
consequence the partition function can be identified with the affine
character of $\widehat{\lambda}$
\be
Z_{\lambda}(v,\tau) = \Tr_{V_{\widehat{\lambda}}}\left(y^{J_0} q^{L_0-c/24} \right) = \chi^{\widehat{\mathfrak g}}_{\widehat{\lambda}}(v,\tau) = \sum_{m,n} d(m,n) y^m q^{n+h_{\widehat{\lambda}} - Nk/24},   \label{Z-VW}
\ee
with $\lambda$ specifying the boundary conditions on boundary of ALE
space $S^3/\Gamma$ and $y=e^{2\pi i v}$. In particular for
$\Gamma=\Z_k$ and the ALE space of $A_{k-1}$ type we get a
representation of $\widehat{su}(k)_N$ and in this case dual Dynkin
indices $d_i=1$ for $i=0,\ldots,k-1$. With the boundary condition
$\lambda$ at infinity we get a vector-valued partition function whose
components are of the form
\be
Z_{\lambda}(v,\tau) = \chi^{\widehat{su}(k)_N}_{\widehat{\lambda}}(v,q).  \label{Z-su-k-N}
\ee
For $U(1)$ the above characters are given explicitly in appendix
\ref{app-characters}. In the following sections we show how these
partition functions are reproduced by some particular classes of
two-dimensional partitions.

Let us also mention, that the McKay correspondence for $A_{k-1}$ ALE
spaces was derived from a string theory perspective in \cite{dhsv} by
relating gauge theory on a Taub-NUT space to the intersecting brane
configuration. Taub-NUT space is a deformation of ALE space which
approaches $S^1$ of constant radius at infinity. In that case there
are additional monopoles around this $S^1$, which is manifested by
additional $\chi^{\widehat{u}_1}$ factors in the partition function
(\ref{Z-su-k-N}).


\section{Orbifold partitions} \label{sec-crystals}

An ordinary two-dimensional partition $\lambda$ can be seen to
determine an ideal of functions $\mathcal{I}=\{f(x,y)\} \subset
\mathbb{C}[x,y]$ generated by a set of monomials $x^i y^j$ for $i,j
\geq 0$, in such a way that a box $(m,n)\in \lambda$ if and only if
$x^m y^n \notin \mathcal{I}$.

\begin{figure}[htb]
\begin{center}
\includegraphics[width=0.25\textwidth]{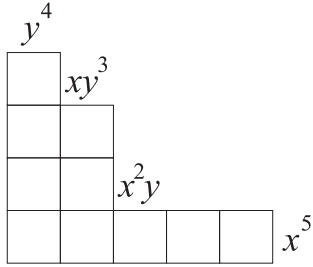}
\begin{quote}
\caption{An ideal $\mathcal{I}\subset \mathbb{C}[x,y]$ generated by the set of monomials $(x^5,x^2y,xy^3,y^4)$ corresponds to the partition $(5,2,2,1)$.}
\end{quote}
\end{center}
\end{figure}

We introduce the following $\mathbb{Z}_k$ orbifold action on $\mathbb{C}^2$
\begin{equation}
(x,y) \to (\omega x,\overline{\omega}y),\qquad \textrm{for}\ \omega=e^{2\pi i /k}.  \label{Zk-action}
\end{equation}

Let us consider ideals of functions having definite transformation
properties under this action. A given monomial $x^i y^j$ transforms as
$\omega^{i-j}$, and all monomials with the same transformation
property can be represented as a periodic sub-lattice of
$\mathbb{Z}^2$. In particular, there is a set of invariant monomials
which we refer to as \emph{invariant} (or \emph{non-twisted}) sector
(these necessarily include constant functions represented by $(0,0)$
point on the lattice). All the other classes of monomials will be
called \emph{twisted} sectors.

We define two types of partitions, which we call \emph{orbifold} or \emph{generalized  partitions}, as follows:
\begin{itemize}
\item {\bf Orbifold partition of the first type:} this is an ordinary 
two-dimensional partition, however with some subset of its boxes distinguished; 
these distingusihed boxes, as points in $\mathbb{Z}^2$ lattice, correspond to monomials 
with a definite transformation property under the action (\ref{Zk-action}); 
we define a weight of such a partition as the number of these distinguished boxes 
(and \emph{not all boxes} in this partition),
\item {\bf Orbifold partition of the second type:} this is a partition whose diagram consists \emph{only} of these distinguished boxes, with a weight given by their number.
\end{itemize}
We draw these distinguished boxes in black in the figures below. In the first case, there are generally many partitions with the same set of distinguished boxes, but with different positions of remaining (``weightless'') boxes. In the second case, a given set of distinguished boxes defines one and only one partition. One can also think of the partitions of the second type as equivalence classes of partitions of the first type, such that all elements in one class have the same set of distinguished boxes. 


\begin{figure}[htb]
\begin{center}
\includegraphics[width=0.8\textwidth]{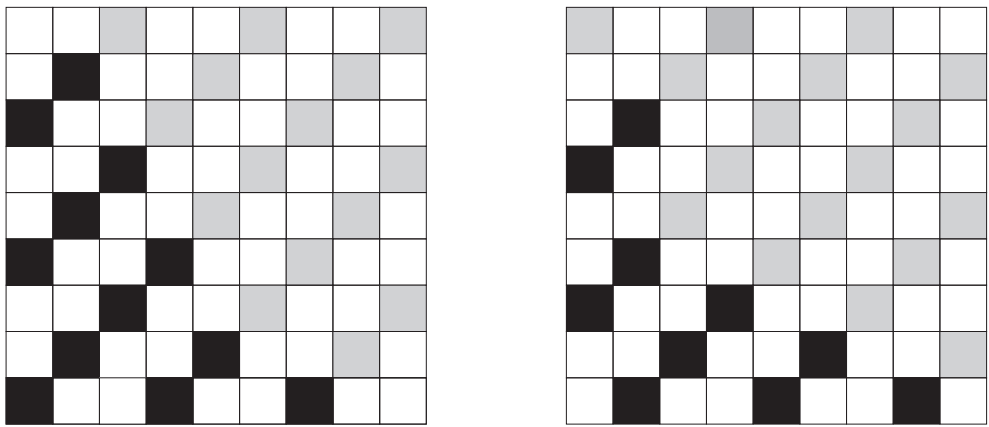}
\begin{quote}
\caption{Sample generalized partitions of the second type (consisting
  only of black boxes) in the invariant (left) and the twisted (right)
  sectors of $\mathbb{C}^2/\mathbb{Z}_{3}$
  orbifold.} 
\end{quote}
\label{fig-orbi-Z3}
\end{center}
\end{figure}

Examples of generalized $\mathbb{Z}_3$-partitions of the second type
are given in figure \ref{fig-orbi-Z3}. The left picture corresponds to
the invariant sector, and the right one to the twisted sector
corresponding to monomials transforming with a factor of
$\omega=\exp(2\pi i/3)$. The corresponding lattices are denoted by
grey boxes; the box in the left-bottom corner has coordinates $(0,0)$
and corresponds to constant functions. Sample partitions in both
sectors are denoted in black, and white boxes are immaterial (examples
of partitions of the first type arise if we take white boxes into
account as well). A proper way to define a generalized partition (of
both types and in any sector) is as follows: if a box $(m,n)$ belongs
to a partition, then all the boxes $(i,j)$ from the (grey) sub-lattice
such that $i\leq m$ and $j \leq n$ must also belong to this partition;
this property is easily seen in the figure \ref{fig-orbi-Z3}.

To work with generalized partitions of the second type it is crucial
to denote them in a way which takes into account only the
distinguished boxes. The usual convention to write the number of
distinguished boxes in each row is not the best choice, as these
numbers are not decreasing; for example the partitions in figure
\ref{fig-orbi-Z3} correspond respectively to sequences
$(3,2,1,2,1,1,1,1)$ and $(3,2,2,1,0,1,1)$. An additional condition
which states when a number of boxes in the next row may increase must
be introduced in this case; this condition is simple but awkward.

\begin{figure}[htb]
\begin{center}
\includegraphics[width=0.5\textwidth]{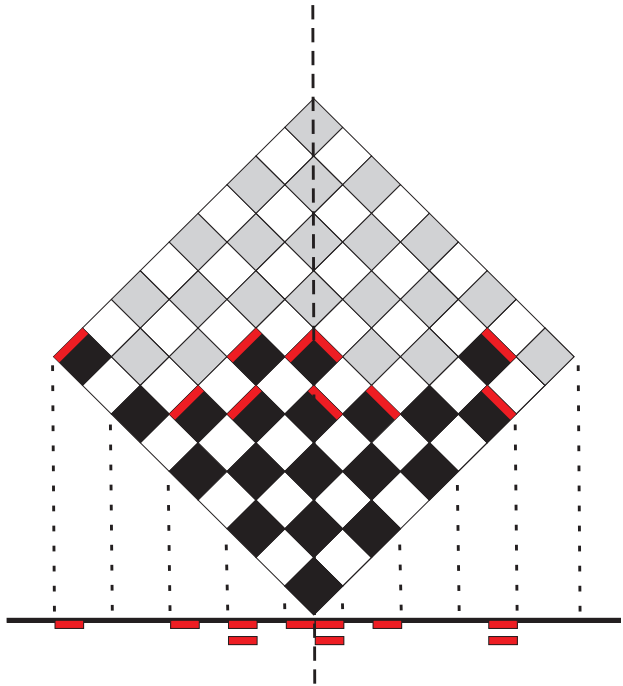}
\begin{quote}
\caption{A generalized partition of the latter type in an invariant
  sector of $\mathbb{C}^2/\mathbb{Z}_{2}$ orbifold as a state in a
  parafermi sea, as projected onto the horizontal axis. Note that
  because of parastatistics states can be at most twice occupied.}
\label{fig-orbi-Z2-fermion}
\end{quote}
\end{center}
\end{figure}

It turns out that a better idea is to use the so-called Frobenius notation
introduced in appendix \ref{app-partitions}. For definiteness, let us
focus on an invariant sector. We slice a given partition $\lambda$
diagonally and introduce two sequences of numbers $(a_i)$ and $(b_j)$,
which denote number of boxes in rows to the right and to the left of
the diagonal
$$
\lambda = \left( \begin{array}{lllll} 
a_1 & a_2 & \ldots & a_{d(R)} \\
b_1 & b_2 & \ldots & b_{d(R)} \\
\end{array} \right)
$$ 
where $d(R)$ is the number of boxes on the diagonal. For ordinary
partitions, the sequences $(a_i)$ and $(b_i)$ must be strictly
decreasing. For generalized $\mathbb{Z}_k$-partitions, a crucial
point is that this condition is relaxed: a given number can
occur at most $k$ times. This coincides precisely with the definition
of \emph{generalized Frobenius partitions}, as was introduced in
\cite{Andrews}. Let us note that such a partition can be presented as
a state of a Fermi sea with a generalized statistics: there may be at
most $k$ fermions at a given position (such objects are called
\emph{parafermions}, and their unusual form of the Pauli principle a
\emph{parastatistics}). In figure \ref{fig-orbi-Z2-fermion} the
partition
$$
\left( \begin{array}{lllll} 
3 & 3 & 1 & 0 & 0 \\
4 & 2 & 1 & 1 & 0 \\
\end{array} \right)
$$ 
is presented for $\mathbb{C}^2/\mathbb{Z}_2$ orbifold, together
with the corresponding state in the parafermi sea. Indeed, in this
case at most two parafermions can sit in the same
place. Generalization of this setup to other sectors is
straightforward.

\section{Counting of partitions of the first type} \label{sec-crystals-resum-0}

In this section we wish to compute generating functions of the generalized partitions of the first type for ALE spaces of $A_{k-1}$ type
$$
\tilde{Z}^k_r = \sum_{\hbox{\scriptsize \it first type partitions}} q^{\# (black\ boxes)},
$$
in all sectors $r=0,\ldots,k-1$, where $r$ specifies the power of $\omega$ in (\ref{Zk-action}).

As we show below, these generalized partitions are related to states
in the Fock space of $k$ free fermions, and the number of black boxes
gives weights of such states. As is well known, a system of $k$ free
fermions provides a representation of $\widehat{u}(k)_1$ affine
Kac-Moody algebra. For this reason $\tilde{Z}^k_r$ can be expressed in
terms of affine characters. On the other hand, the black boxes
transform in a definite way under the $\mathbb{Z}_k$ orbifold group of
$A_{k-1}$ ALE space, which provides a relation to Nakajima's results.

We start with the observation that generalized partitions of the first
type can be identified with \emph{blended} partitions, which describe
a state of a Fermi sea of several fermions. Let us consider $k$
complex fermions with corresponding fermion charges $p_1,\ldots,p_k$, and fixed total charge
$p$
$$
p=\sum_{i=1}^{k} p_i.
$$ 
For the $i^{th}$ fermion there is a corresponding Fock space
$\mathcal{F}_i$, and elements of its basis can be represented in a
standard way as states in a Fermi sea, or in terms of the usual Young
diagrams with a specified charge $p_i$. The total Fock space
$\mathcal{F}$ is a tensor product of $k$ of these Fock spaces 
$$
\mathcal{F} = \bigotimes_{i=1}^{k} \mathcal{F}_i.
$$ The basis elements of $\mathcal{F}$ are obtained by tensoring the
basis elements of $\mathcal{F}_i$. Tensor products of states that
correspond to a colored partition $\vec{{\bf R}}=\{ R_{(i)} \}$ with
charges $p_i$, are also in one-to-one correspondence with a particular
kind of two-dimensional partitions, which are called blended
partitions \cite{Nek-Ok,jimbo-miwa}.
A blended partition ${\bf R}=( {\bf R}_K )_{K\in
  \mathbb{N}}$ is defined by a set of (necessarily distinct) integers
\be \{ k(p_i+R_{(i),m} - m) +i-1\ |\ m\in\mathbb{N} \} = \{ p+{\bf
  R}_K -K \ | \ K\in\mathbb{N} \}, \label{blend} \ee which implicitly
defines a finite number of non-zero ${\bf R}_K$ ordered such that
${\bf R}_1 \geq {\bf R}_2 \geq \ldots$. The total number of boxes of
such a partition is equal to \be | {\bf R} | = \sum_i \Big( k|R_{(i)}|
+ \frac{k}{2} p_i^2 + ip_i \Big) - \frac{(k+1)p}{2} -
\frac{p^2}{2}. \label{blend-nr} \ee Extending the relation between
partitions and chiral fermions explained in appendix
\ref{app-fermion}, one can of course write the definition of a blended
partition equivalently in terms of fermionic states, by merging $k$
chiral fermions into a single chiral fermion.

\begin{figure}[htb]
\begin{center}
\includegraphics[width=0.8\textwidth]{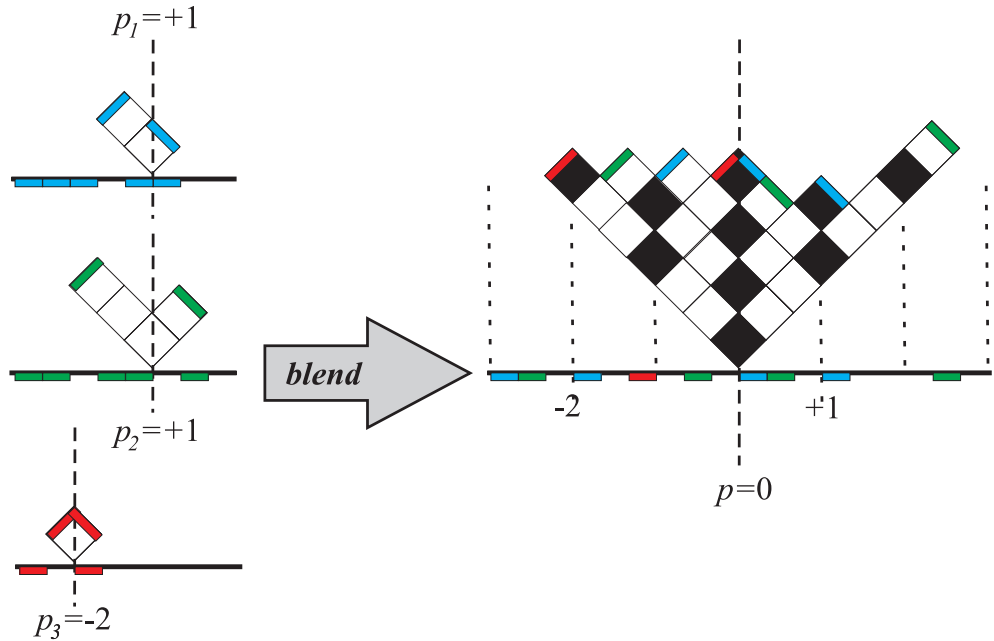}
\begin{quote}
\caption{Blended partition obtained from $k=3$ fermions is in fact equivalent to generalized partition. The total fermion scaling weight (energy) $\sum(|R_{(i)}| + p_i^2/2) = 10$ is determined by the number of black boxes.} 
\label{fig-orbi-Z3-blend}
\end{quote}
\end{center}
\end{figure}

Now we claim the generalized $\mathbb{Z}_k$-partitions of the first type are in one-to-one correspondence with blended partitions ${\bf R}$ obtained from $k$-colored partitions $\vec{{\bf R}}$, such that
\begin{itemize}
\item a generalized partition has the same shape as the corresponding blended partition,
\item a weight of a generalized partition (as given by the number of distinguished boxes it contains) is specified by the total weight of a state of $k$ fermions related to $\vec{{\bf R}}$. 
\end{itemize}
A generalized partition corresponding to a certain blended one is shown in figure \ref{fig-orbi-Z3-blend}. 

\begin{figure}[htb]
\begin{center}
\includegraphics[width=0.6\textwidth]{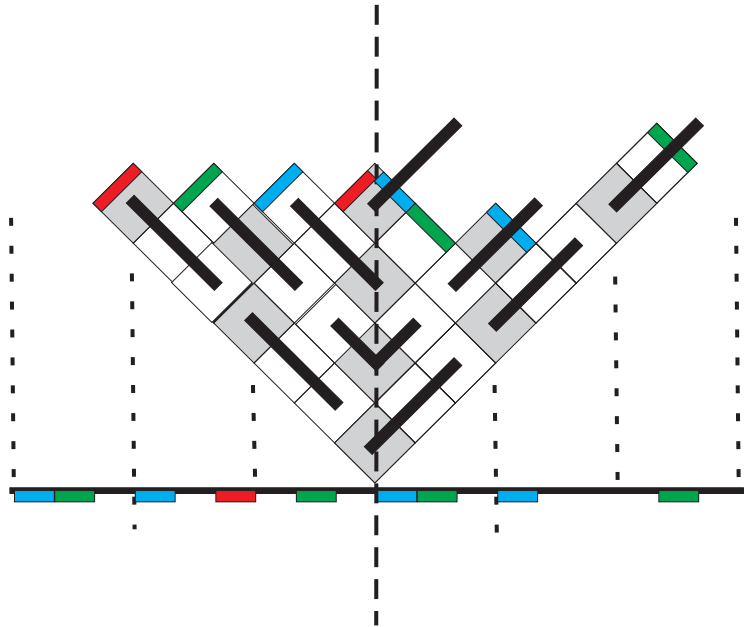}
\begin{quote}
\caption{Counting of distinguished boxes of a generalized partition
  from figure \ref{fig-orbi-Z3-blend}. This partition consists of 4
  hooks attached to the diagonal. The number of ``excess'' boxes is
  $-1-1+1-2=-3$, which can be computed following
  eqn. (\ref{excessive}) as $\sum_i ip_i = 1+2*1+3*(-2)$ (in this case
  $n=r=0$).}
\label{fig-blend-inv}
\end{quote}
\end{center}
\end{figure}

The total weight of a state of  $k$ fermions is equal to their contribution to the character (\ref{u-k-1})
\be
\sum_i \Big( |R_{(i)}| + \frac{p_i^2}{2} \Big).  \label{black-boxes}
\ee
To find the number of distinguished boxes in a generalized partition
${\bf R}$ (of the same shape as a blended partition) it is helpful to
represent it as a set of hook partitions (having precisely a single
row and a single column), fixed to diagonal elements of ${\bf R}$, and
divided into intervals of at most $k$ boxes. An example of such a
construction for a situation in figure \ref{fig-orbi-Z3-blend} is
given in figure \ref{fig-blend-inv}, with these intervals drawn in
black. Each interval of $k$ boxes contains precisely one distinguished
box. To count distinguished boxes, we will count all boxes contained
in those black intervals and divide their number by $k$. However,
there are some ``excess'' boxes: in fact some of those black
intervals contain less than $k$ boxes (when an interval sticks out of
the partition); or some white boxes don't belong to any black interval
(when all distinguished boxes in a given hook have already been
matched to some black interval). Let us rewrite the charges $p_i$ in terms
of $n,r,n_1,\ldots,n_{k-1}$ as in (\ref{p-decompose}), which is always
possible in a unique way. In particular, we have
$$
n = [p/k],\qquad r=p-kn,
$$
where $[\cdot]$ denotes the integer part of a real number. Let us also
introduce
$$
p'_i = p_i - n -\delta,\qquad \textrm{where}\quad \delta=\left\{
\begin{array}{cl}
1 & \textrm{for}\ i=1,\ldots,r \\
0 & \textrm{for}\ i=r+1,\ldots,k
\end{array}\right.
$$
Now it is straightforward to show that the number of these excessive boxes is given by
\be
\sum_i ip'_i = \sum_i ip_i -\frac{(k+1)k n}{2} - \frac{(r+1)r}{2}.  \label{excessive}
\ee
Subtracting this from the total number of boxes $| {\bf R}|$ and dividing by
$k$ we conclude the number of distinguished boxes is equal to
$$
\frac{|{\bf R}| - \sum_i ip'_i}{k} = \sum_i \big( |R_{(i)}| + n_i^2 - n_i
n_{i+1}  \big) + \frac{r^2}{2} + n_1 r - \frac{r}{2},
$$
which almost reproduces (\ref{black-boxes}), so that summing over all diagrams $R_{(i)}$ with the total fixed charge $p=kn+r$ we get
\bea
\tilde{Z}^k_r & = & \sum_{1^{st}\ type\ partitions} q^{\# (black\ boxes)} = \frac{q^{k/24}}{\eta(q)^k}
\sum_{n_1,\ldots,n_{k-1}} q^{\sum_i (n_i^2 - n_i n_{i+1}) + \frac{r^2}{2} +
  n_1 r - \frac{r}{2}} = \nonumber \\
& = & \frac{q^{\frac{k}{24} + \frac{r^2}{2k} - \frac{r}{2}}}{\eta(q)} \chi^{\widehat{su}(k)_1}_r(0),  \label{Z1-k-r}
\eea
which is proportional to the $\widehat{su}(k)_1$ characters as given by equation (\ref{su-char}) in Appendix C, as
computed for $z_i=0$. Now following the formulae of this Appendix, we note that the $\widehat{u}(k)_1$ character decomposes according to equation (\ref{char-u-su}) into $k$ level 1 affine characters indexed by $r=0,\ldots,k-1$, weighted by
$\widehat{u}(1)_k$ characters
\begin{equation}
\chi^{\widehat{u}(k)_1}(x_i) = \sum_{r=0}^{k-1} \chi^{\widehat{u}(1)_k}_r(\tilde{x})\, \chi^{\widehat{su}(k)_1}_r(\tilde{x}_i).
\end{equation}
Here $x_i$, $i=1,\ldots,k$ are the specialization points given in equation
(\ref{char-u-vars}), which in particular determine variables $y_j$,
$j=1\ldots k-1$ in which $\widehat{su}(k)_1$ characters are naturally
expressed, see equation (\ref{char-su-vars}). We see that generating functions for
generalized partitions indeed combine into $\widehat{u}(k)_1$
character computed at values of $x_i=\exp(2\pi i z_i) = 1$. Summing
(\ref{Z1-k-r}) over allowed $r$ we reproduce the $n=0$ sector of
$\widehat{u}(k)_1$ character 
\be
\chi^{\widehat{u}(k)_1}(z_i=0)|_{n=0} = q^{-k/24} \sum _{r=0}^{k-1}
q^{r/2}\,\tilde{Z}^k_r.  \label{Z1-uk-chi}
\ee
Thus the counting of states of $k$ fermions with fixed total charge is
equivalent to the counting of generalized partitions of the first type. 

Let us finally note that we always can think of the distinguished
boxes we count as belonging to the \emph{invariant} sector, {\it i.e.}
given by monomials $x^i y^j$ that are invariant under the $\mathbb{Z}_k$
action. If $p$ is a multiplicity of $k$, the box at position $(0,0)$
of a blended partition always transforms invariantly. But if the total
charge $p$ is not a multiplicity of $k$, the corner of the blended
partition is fixed at the position $p$ of the total Fermi sea, and the
box $(0,0)$ does not transform invariantly; thus those boxes which do
transform invariantly may be thought of as belonging to twisted
sectors of the generalized partition.

\section{Counting of partitions of the second type} \label{sec-crystals-resum}

Now we consider generalized partitions of the second type. We prove their
generating functions are also given by affine characters, this time computed at
some special values of parameters $y_i$. Thus they can be identified with equivalence classes of blended partitions mentioned above.

To start with, for a given $k$ we consider the following function of two parameters introduced in \cite{Andrews}
\begin{eqnarray}
G_k(z,q) & = & \prod_{n=0}^{\infty} (1+z q^{n+1} +\ldots z^k q^{k(n+1)})(1+z^{-1}q^n +\ldots z^{-k} q^{k n} ) \nonumber \\
& = & \sum_{r\in\mathbb{Z}} Z^{k}_r(q) z^{r},  \label{G-N}
\end{eqnarray}
where the second line defines implicitly functions $Z^{k}_r(q)$.

The crucial observation is that this expression encodes generating
functions for all non-twisted and twisted sectors of
$\mathbb{C}^2/\mathbb{Z}_{k}$ orbifold: a partition function for a
sector twisted by $\omega^r$ (for $r=0,\ldots,k-1$) is given by
$Z^{k}_{-r}(q)$, i.e. a term proportional to $z^{-r}$. It is
understood that the invariant sector corresponds to $i=0$,
i.e. $z$-independent term. Thus we claim
$$
Z^k_{-r} = \sum_{\hbox{\scriptsize \it second type partitions}} q^{\# (black\ boxes)}.
$$ 
We recall the generalized partitions can be interpreted as blended partitions, and then there is always only one type of distinguished boxes, related to invariant monomials from the total Fock space $\mathcal{F}$ point of view. But for $r\neq 0$ the corner of this blended partition is fixed at such a position that these distinguished invariant boxes belong to the twisted sector of the partition.

In particular, for $k=1$ the expression (\ref{G-N}) reduces to the
standard Jacobi triple product identity (\ref{Jacobi}), with a single
sector with $Z^{k=1}_{0}(q) \sim \prod (1-q^n)^{-1} = q^{-1/24}
\eta(q)$. This reproduces of course the partition function for
$\mathcal{N}=4$ theory on $\mathbb{R}^4$, given by $\eta(q)$
function. We now present how it generalizes to ALE spaces of $A_{k-1}$
type for arbitrary $k$.

As mentioned above, our generalized partitions from the invariant
sector of $\mathbb{C}^2/\mathbb{Z}_k$ lattice are in one-to-one
correspondence with generalized Frobenius partitions with $k$
repetitions allowed. Their generating function was shown in
\cite{Andrews} to be given by $Z^k_0(q)$. We extend now this
observation to all other sectors in order to include instanton
contributions for all possible flat connections at infinity.

\begin{figure}[htb]
\begin{center}
\includegraphics[width=0.6\textwidth]{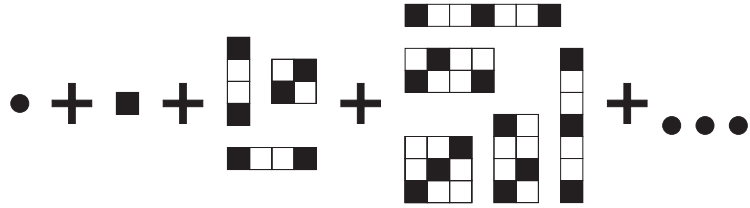}
\begin{quote}
\caption{Orbifold partitions of the second type in an invariant sector consist of black boxes which are invariant with respect to the action of the orbifold group. First terms of a generating function for an invariant sector of $\mathbb{C}^2/\mathbb{Z}_{3}$ orbifold are $Z^3_0=1+q+3q^2+5q^3+\ldots$} 
\label{fig-orbi-Z3-sum}
\end{quote}
\end{center}
\end{figure}

\begin{figure}[htb]
\begin{center}
\includegraphics[width=0.6\textwidth]{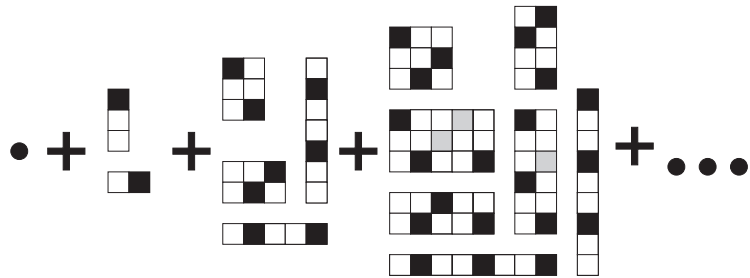}
\begin{quote}
\caption{First terms of the generating function of the twisted sector of 
the $\mathbb{C}^2/\mathbb{Z}_{3}$ orbifold $Z^3_{-1}=Z^3_{-2}=1+2q+4q^2+7q^3+\ldots$} 
\label{fig-orbi-Z3-sum-twist}
\end{quote}
\end{center}
\end{figure}

Thus, our present aim is to compute generating functions $Z^{k}_{-r}(q)$ for each $r=0,\ldots,k-1$. A symmetry under reflection along the diagonal $x=y$ implies the relation
$$
Z^k_{-r}=Z^k_{-(k-r)}.
$$
Anticipating the result, this is also an important property of $\widehat{su}(k)$ affine characters at level 1. Several lowest terms in generating functions for invariant and twisted sectors of $\mathbb{C}^2/\mathbb{Z}_{3}$ orbifold, together with all relevant partitions, are shown in figures \ref{fig-orbi-Z3-sum} and \ref{fig-orbi-Z3-sum-twist}.

The function (\ref{G-N}) can be rewritten as follows
\begin{eqnarray}
G_k(z,q) & = & \prod_{j=1}^{k} \Big(\prod_{m=1}^{\infty}(1-\zeta^j z q^m)\,\prod_{n=0}^{\infty}(1-\zeta^{-j} z^{-1} q^n) \Big) = \label{G-fermions} \\
& = & \prod_{i=1} \frac{1}{(1-q^i)^k} \prod_{j=1}^{k} \sum_{m_j\in \mathbb{Z}} (-1)^{m_j} q^{m_j(m_j+1)/2} z^{m_j} \zeta^{j m_j} \nonumber
\end{eqnarray}
where $\zeta = e^{2\pi i / (k+1)}$. (Note it's different from $\omega$!)

The first line above reveals a relation to a system of $k$ fermions with phases $\zeta^{\pm j}$. The second line allows to extract a term $Z^k_{-r}(q)$ we are looking for by imposing the condition
\begin{equation}
-r = m_1+m_2+\ldots + m_k.
\end{equation}
After a little  algebra this leads to the main result:
\begin{equation}
Z^k_{-r} = (-1)^r q^{r(r-1)/2} \zeta^{- k r} \frac{q^{k/24}}{\eta(q)^k} \times \label{Z-N-k}
\end{equation}
$$ 
\times \sum_{m_1,\ldots,m_{k-1}\in\mathbb{Z}} q^{\sum_{i=1}^{k-1} m_i^2  +  \sum_{i<j} m_i m_j + r\sum_{i=1}^{k-1} m_i}  \zeta^{-(k-1)m_1 - (k-2)m_2 -\ldots -m_{k-1}},
$$
where $r=0,\ldots,k-1$.

Let us now prove that the generating functions we obtained in
(\ref{Z-N-k}) indeed can be written as affine $\widehat{su}(k)_1$
characters, and moreover that they encode the entire $\widehat{u}(k)_1$
character, similarly as in the first case (\ref{Z1-uk-chi}).

First, let us redefine the summations in (\ref{su-char}) by introducing $n_i = \sum_j M_{i j}m_j$, for $M$ having $1$ on and above the anti-diagonal (and zeros otherwise). We get
$$
\chi^{\widehat{su}(k)_1}_r(y_i) = \frac{1}{\eta(q)^{k-1}} q^{\frac{r^2}{2}\frac{k-1}{k}}\zeta^{-r(k-1)/2} \times
$$
$$    
\times \sum_{m_1,\ldots,m_{k-1}} q^{\sum_i m_i^2 + \sum_{i<j} m_i m_j + r\sum m_i} \zeta^{-m_1(k-1)\ldots -m_{k-1}},
$$
and to match to $Z^k_{-r}$ in (\ref{Z-N-k}) we had to choose 
$$
y_i = \zeta^{a_i},\quad \textrm{for}\ a_i = -\sum_j A^{-1}_{ij} = -\frac{(k-i)i}{2},
$$
which immediately determine the ratios of $x_i$'s in (\ref{char-u-vars})
$$
y_i = \zeta^{a_i} \iff \tilde{y}_i = \frac{x_i}{x_{i+1}} = \zeta^{-1}.
$$
Moreover, the prefactors in (\ref{Z-N-k}) match the $n=0$ factor of the $\widehat{u}(1)_k$ character if we identify
$$
\tilde{x} = x_1\cdots x_k = \zeta^{-k(k+1)/2},
$$
which altogether determines
$$
x_i = \zeta^{-k-1+i} = \zeta^{i},
$$
where we used $\zeta^{k+1}=1$.

Now the precise relation between the characters and the orbifold partitions reads
\be
Z^k_{-r} = \sum_{\hbox{\scriptsize \it second type partitions}} 
q^{\# (black\ boxes)} = f_{k,r}(q,\zeta)\,\chi^{\widehat{su}(k)_1}_r(y_i=\zeta^{a_i}), \label{Z-chi}
\ee
where
$$
f_{k,r} =  (-1)^r\,\zeta^{-\frac{r}{2}-\frac{rk}{2}}\, \frac{q^{\frac{k}{24}+\frac{r^2}{2k}-\frac{r}{2}}}{\eta(q)} = (-1)^{r} q^{\frac{k}{24}-\frac{r}{2}}\,\chi^{\widehat{u}(1)_k}_r \big(\tilde{x}=\zeta^{-\frac{k(k+1)}{2}}  \big)|_{n=0},
$$
where $|_{n=0}$ denotes the corresponding term in equation (\ref{u1-char}). 

Finally, the full partition function arises from different sectors and should be given by the overall $\widehat{u}(k)_1$ character as in (\ref{char-u-su}). From the above identifications we get
\be
\chi^{\widehat{u}(k)_1}(x_i=\zeta^i)|_{n=0} = q^{-k/24} \sum_{r=0}^{k-1} (-1)^r q^{r/2} \, Z^k_{-r}.  \label{Z-uk-chi}
\ee


\section{Some examples of orbifold partitions of the second type}  \label{sec-examples}

The generating functions we obtained can also be presented in terms of infinite products by applying the Jacobi identity
\begin{equation}
\prod_{n=1}^{\infty} (1-q^n)(1+z q^n)(1+z^{-1}q^{n-1}) = \sum_{m \in \mathbb{Z}}z^m q^{m(m+1)/2}.  \label{Jacobi}
\end{equation}
Some examples of particular computations corresponding to formula (\ref{Z-N-k}) for the partitions of the second type for orbifolds $\mathbb{C}^2/\mathbb{Z}_k$ for $k=2,3,4$ are given below.

\subsection{$\mathbb{C}^2/\mathbb{Z}_2$}

The invariant sector:
\begin{eqnarray}
Z^{2}_0(q) & = & \prod_{n=1} \frac{1}{(1-q^n)(1-q^{12n-10})(1-q^{12n-9})(1-q^{12n-3})(1-q^{12n-2})} = \nonumber \\
& = & \prod_{n=1} \frac{1}{1-q^n} \exp \sum_{n>0} \frac{q^{-4n}+q^{-3n}+q^{3n}+q^{4n}}{(-1)\cdot n[12n]} = \\
& = & 1+q+3q^2+5q^3+9q^4+14q^5+24q^6+35q^7+55q^8+81q^9+\dots \nonumber
\end{eqnarray}
The twisted sector (terms proportional to $\omega =e^{2\pi i /2} =-1$):
\begin{eqnarray}
Z^{2}_{-1}(q) & = & \prod_{n=1} \frac{1+q^{2n-1}}{(1-q^n)(1-q^{12n-6})} = \\
& = & \Big( \prod_{n=1} \frac{1}{1-q^n} \Big)\  \exp \sum_{n>0} \frac{q^{-5n}+q^{-3n}+q^{-n}-(-1)^n+q^n+q^{3n}+q^{5n}}{(-1)^n\cdot n[12n]} = \nonumber \\
& = & 1+2q+3q^2+6q^3+10q^4+16q^5+26q^6+40q^7+60q^8+90q^9+\dots \nonumber
\end{eqnarray}

\subsection{$\mathbb{C}^2/\mathbb{Z}_3$}

The invariant sector (compare with figure \ref{fig-orbi-Z3-sum}):
\begin{eqnarray}
Z^{3}_0(q) & = & \prod_{n=1} \frac{(1-q^{6n})(1-q^{12n-6})}{(1-q^n)(1-q^{2n})(1+q^{6n})(1-q^{6n-3})^2} = \nonumber \\
& = & \prod_{n=1} \frac{1}{(1-q^n)} \\
& &  \exp \sum_{n>0} \frac{q^{-4n}+2q^{-3n}+q^{-2n}-1+(-1)^n+q^{2n}+2q^{3n}+q^{4n}+(-1)^nq^{6n}}{(-1)\cdot n[12n]} = \nonumber \\
& = & 1+q+3q^2+6q^3+11q^4+18q^5+31q^6+49q^7+78q^8+119q^9+\dots \nonumber
\end{eqnarray}

Both twisted sectors (terms proportional to $\omega=e^{2\pi i /3}$ and $\overline{\omega}$) have the same generating functions, as is geometrically obvious; this function in fact is a sum of two infinite products (compare with figure \ref{fig-orbi-Z3-sum-twist}):
\begin{eqnarray}
Z^{3}_{-1;-2}(q) & = & -i \prod_{n=1} \frac{(1-q^{6n})(1-q^{2n})}{(1-q^n)^3}
\Big(\prod_{n=1} (1+iq^{6n-1})(1-iq^{6n-5})(1+q^{4n-2}) + \nonumber \\
& & +(i-1)\prod_{n=1} (1+iq^{6n-4})(1-iq^{6n-2})(1+q^{4n})   \Big) \\
& = & 1+2q+4q^2+7q^3+13q^4+22q^5+36q^6+57q^7+90q^8+137q^9+\dots \nonumber
\end{eqnarray}

\subsection{$\mathbb{C}^2/\mathbb{Z}_4$} \label{sss-C2Z4}

We present just several terms in the expansion. For the invariant sector:
$$
Z^{4}_0(q) = 1+q+3q^2+6q^3+12q^4+20q^5+35q^6+56q^7+92q^8+142q^9+\dots
$$

Twisted sectors proportional to $\omega=e^{2\pi i /4}=i$ and $\overline{\omega}=-i$ have the same generating functions:
$$
Z^{4}_{-1;-3}(q) =  1+2q+4q^2+8q^3+14q^4+25q^5+42q^6+68q^7+108q^8+168q^9+\dots
$$

The twisted sector proportional to $\omega^2=-1$:
$$
Z^{4}_{-2}(q) = 1+2q+5q^2+8q^3+16q^4+26q^5+45q^6+72q^7+115q^8+176q^9+\dots
$$

\subsection{$\mathbb{C}^2/\mathbb{Z}_4$ and affine characters}

Let us consider an example of a relation between orbifold partitions and characters for the case of $\mathbb{C}^2/\mathbb{Z}_4$. Using
$$
\zeta=e^{\frac{2\pi i}{5}},\qquad \tilde{y}_1=\tilde{y}_2=\tilde{y}_3 = \zeta^{-1},\qquad x_i = \zeta^i
$$
in formulas (\ref{Z-chi}) and (\ref{su-char}), one can immediately rederive expansions in section \ref{sss-C2Z4}. Moreover, using these expansions, the decomposition of $\widehat{u}(4)_1$ character (\ref{u-k-1}) indeed matches the result (\ref{Z-uk-chi})
$$
\chi^{\widehat{u}(4)_1}(x_i=\zeta^i)|_{n=0} = q^{-1/6}\Big( Z^4_{0} - q^{1/2}Z^4_{-1}+qZ^4_{-2}-q^{3/2}Z^4_{-3} \Big) =
$$
$$
 = q^{-1/6} \big( 1 -q^{1/2}+2q-3q^{3/2}+5q^2-6q^{5/2}+11q^3-12q^{7/2}+20q^4-22q^{9/2}+36q^5 \big).
$$


\section{Summary} \label{summarize}

In this paper we found an interpretation of partition functions of
$\mathcal{N}=4$ theories on ALE spaces in terms of generalized (or
orbifold) partitions, and we described the corresponding free fermion
system. In particular, we obtained two expressions for the partition
function of the $\mathcal{N}=4$ theory given by equations
(\ref{Z1-uk-chi}) and (\ref{Z-uk-chi}). These expressions encode
$\widehat{su}(k)_1$ characters which is consistent with predictions of
\cite{naka1,V-W}, and can also be viewed as $n=0$ sector of $\widehat{u}(k)_1$
characters. Summation over all $n$ would reproduce the full $\widehat{u}(1)$
factor of $\widehat{u}(k)_1$ characters, which was interpreted in \cite{dhsv} 
as arising from monopoles going around $S^1$ at infinity of the Taub-NUT
space (this Taub-NUT is a circle compactification of ALE
space and can be used to engineer the relevant
$\mathcal{N}=4$ theory dual to our fermionic system by 9-11 duality). As
$\widehat{u}(k)_1$
characters expressions (\ref{Z1-uk-chi}) and (\ref{Z-uk-chi}) differ only in
their arguments, and 
both can be understood as generating functions of generalized
partitions. These generalized partitions are identified with blended
partitions representing $k$ fermions with charges $p_i$ and fixed
total charge $p=\sum_i p_i$. In the character the fermions are
weighted by a power of $q$ which turns out to be equal to the number
of distinguished (black) boxes in corresponding generalized
partition. In the first case each blended partition arises once, so in
general there are several partitions with the same set of black boxes,
but differing in positions of white ``weightless'' boxes. In the
latter case the effect of a particular value of $x_i=\zeta^i$ is such
that coefficients of all generalized partitions with the same
configuration of black boxes add up exactly to 1, so the counting
reduces to the counting of the generalized Frobenius partitions
introduced by Andrews.

It would be nice to generalize the point of view we presented in this paper in various directions.
Firstly, it would be interesting to introduce orbifold partitions for other
types of ALE spaces on one hand, and other gauge groups on the other
hand. Moreover, we note that in \cite{Nek-Ok} $U(N)$ gauge theories on $\R^4$
were related to the counting of blended partitions, while our results relate
blended partitions to $U(1)$ theory on ALE spaces. Invoking level-rank
duality, one might hope to connect blended partitions, and possibly more
general orbifold partitions, to $U(N)$ theories on arbitrary ALE spaces, with
$\widehat{su}(k)_N$ characters of Nakajima arising as generating functions of those partitions. 

Secondly, the fermions we considered are related to the fermions living on the
intersection of D4 and D6-branes analyzed in \cite{dhsv}. It would therefore
be of interest to provide an explicit interpretation, in terms of yet more
general class of partitions, of the entire intersecting brane system and its
aspects related to the level-rank duality.

Moreover, our results might be of interest in the context of a Langlands reduction of
four-dimensional gauge theories defined on spaces which are not products of
two curves. One of the simplest examples are $I_k$ singularities, where in the
$I_1$ case instantons of the sigma-model give rise to $\eta$ function as
well. Presumably analogous counting for arbitrary $I_k$ could also be related
to blended partitions.

Finally, one might hope to lift our considerations to three-dimensional partitions and six-dimensional gauge theories. In the simplest case of $\Z_k$ orbifolds, in the first instance it would be desirable to generalize $U(1)$
gauge theory results of \cite{foam} to theories defined on orbifold
$\C^3/\Z_k$ spaces. Even though the corresponding three-dimensional orbifold
partitions are easy to visualize, it is non-trivial to find exactly their
generating functions. Supposedly they are related in a non-trivial way to the
topological vertex theory and topological strings, in a complementary way 
to ensembles of restricted Calabi-Yau crystals considered in \cite{okuda,cube,jafferis}.
This still remains an unexplored area of research.


\bigskip

\bigskip

\centerline{\Large{\bf Acknowledgments}}

\bigskip

We would like to thank Lotte Hollands, Kareljan Schoutens and Cumrun Vafa for discussions
and insightful comments. 

Our research was supported by the FOM programme ``String Theory and
Quantum Gravity'' and a NWO Spinoza Grant. The research of P.S. was
also supported by MNiSW grant N202-004-31/0060.

\bigskip


\appendix

\centerline{\Large{\bf Appendices}}

\section{Two-dimensional partitions} \label{app-partitions}

A \emph{partition} $R=(R_1,R_2,\ldots,R_l)$ is a set of non-increasing positive integers, $R_1\geq R_2\geq\ldots\geq R_l$. $l=l(R)$ is called a \emph{length} of a partition $R$. A partition can be presented in a form of a Young diagram, which is a tableaux of $l(R)$ rows of boxes, with $R_i$ boxes in $i$'th row. We often identify a partition with its Young diagram. A \emph{dual} (or \emph{transposed}) partition arises from a transposition of the Young diagram corresponding to $R$ and is denoted $R^t$.


A partition can also be presented in the so-called Frobenius notation. Let $d(R)$ denote the number of boxes on a diagonal of a Young diagram of $R$. Then
\be
R = \left( \begin{array}{cccc} a_1 & a_2 & \ldots & a_{d(R)} \\
b_1 & b_2 & \ldots & b_{d(R)} \\
\end{array} \right)  \label{frobenius}
\ee
where $a_i$ and $b_i$ have interpretation as distances from each diagonal element to the end of its row and column respectively and are given by
\be
a_i = R_i-i,\qquad b_i = R_i^t-i.  \label{frobenius-ab}
\ee
Sequences $(a_i)$ and $(b_i)$ are necessarily strictly decreasing. 

A size of $R$ is defined as the number of boxes in the corresponding Young diagram. It is denoted $|R|$ and can be written in the standard and the Frobenius notation respectively as
$$
|R|=\sum_{i=1}^{l(R)} R_i = d(R) + \sum_{i=1}^{d(R)}(a_i+b_i).
$$

For example, for a partition
$$
\Yvcentermath1 R=(5,4,2,1,1)=
\left( \begin{array}{cc} 4 & 2  \\
4 & 1  \\
\end{array} \right)
=\yng(1,1,2,4,5)
$$
we have $l(R)=5, d(R)=2,|R|=13$ and 
$$
\Yvcentermath1  R^t = (5,3,2,2,1) =
\left( \begin{array}{cc} 4 & 1  \\
4 & 2  \\
\end{array} \right)
=\yng(1,2,2,3,5)
$$

A \emph{$N$-colored partition} is a set of $N$ partitions $\vec{\bf{R}} = (R_{(1)},R_{(2)},\ldots,R_{(N)})$; each $R_{(i)}$ by itself is a usual partition: $R_{(i),1} \geq R_{(i),2} \geq \ldots \geq R_{(i),l(R_{i})}$. A diagram of $N$-colored partition is a set of $N$ diagrams corresponding to $R_{(i)}$'s. A size of $N$-colored partition $\vec{\bf{R}} = \sum_{i,j} R_{(i),j} $ is equal to the total number of boxes in its diagram.

In this paper we also generalize the notion of a partition and introduce the so-called \emph{orbifold} or \emph{generalized partitions}. As these object are non-standard, we present them in the main body of the paper in section \ref{sec-crystals}. 


\section{Free fermion formalism}   \label{app-fermion}

Let us consider a complex fermion in the NS sector
\be
\psi(z) = \sum_{n\in\Z} \psi_{n+\hf} z^{-n-1},\qquad \psi^*(z) = \sum_{n\in\Z} \psi^*_{n+\hf} z^{-n-1},   \label{fermion-NS}
\ee
subject to anticommutation rules
$$
\{\psi_{n+\hf},\psi^*_{-m-\hf}\} = \delta_{m,n}.
$$
Particle annihilation and creation operators are $\psi^*_{n+\hf}$ with respectively $n\geq0$ and $n<0$. The vacuum state $|0\rangle$ is defined as
\be
\psi_{n+\hf} |0\rangle = 0,\qquad \psi^*_{n+\hf}|0\rangle = 0,\qquad \textrm{for}\ n\geq 0,  \label{fermion-vac}
\ee
and a basis of the total Fock space $\cF$ is given by states obtained by acting with creation operators on this vacuum. The space $\cF$ decomposes 
$$
\cF = \bigotimes_{p\in\Z}\, \cF_p
$$
into subspaces $\cF_p$ of fixed $U(1)$ charge with respect to the current
$$
J(z) = :\psi(z)\psi^*(z): = \sum_{k\in\Z} z^{-k-1} \sum_{n\in\Z} :\psi_{n+\hf}\psi^*_{k-n-\hf}: = \sum_{k\in\Z} J_k z^{-k-1}
$$
and vacua $|p\rangle$ with charge $p$ can be introduced
\bea
\psi_{n+\hf} |p\rangle & = & 0, \qquad \textrm{for}\ n\geq p, \nonumber \\
\psi^*_{m+\hf} |p\rangle & = & 0, \qquad \textrm{for} \ n\geq -p,
\eea
so that each subspace $\cF_p$ is generated from $|p\rangle$. 

There is a very interesting one-to-one correspondence between free fermion states and two-dimensional partitions. In $p=0$ sector the state
$$
|R\rangle = \prod_{i=1}^{d} \psi^*_{-a_i-\hf} \psi_{-b_i-\hf}|0\rangle
$$
corresponds to the partition $R=(R_1,\ldots,R_l)$ such that
$$
a_i = R_i-i,\qquad b_i = R_i^t-i,
$$
which are precisely the numbers which specify a partition in the Frobenius notation (\ref{frobenius}).

\begin{figure}[htb]
\begin{center}
\includegraphics[width=0.6\textwidth]{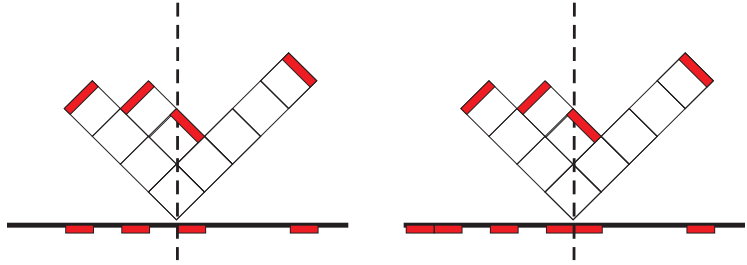}
\begin{quote}
\caption{A correspondence between partitions and states. Positions of particles and holes are given by projecting the ends of the rows and the columns of a partition onto a Fermi sea. Particles are drawn in red, and according to one of the conventions used holes are drawn either in red (left), or just as holes (right). The partition drawn in the figure is $R=(5,2,2,1)$ and the corresponding state is $|R\rangle = \psi^*_{-\hf} \psi_{-\frac{3}{2}} \psi^*_{-\frac{9}{2}}\psi_{-\frac{7}{2}} |0\rangle$.} \label{fig-fermi-part}
\end{quote}
\end{center}
\end{figure}

It is easy to visualize this correspondence in terms of the Fermi sea. In particular the vacuum $|0\rangle$ is given by a Fermi sea with all negative states filled and it is mapped to the trivial partition $\bullet$. A nontrivial partition is most easily visualized if one draws it with a corner fixed at the edge of the filled part of the Fermi sea. Then, the positions of particles and holes are read off by projecting the ends of the rows and the columns of this partition onto the Fermi sea. There are then two conventions to draw such a state of the Fermi sea, as illustrated in figure \ref{fig-fermi-part}. 

For $p\neq 0$ one draws a partition with a corner fixed at position $p$ of the Fermi sea, and then reads off the positions of the particles and the holes similarly as above, by projecting the ends of the rows and the columns onto the Fermi sea. A state corresponding to a partition $R$ of charge $p$ is denoted as $|p,R\rangle$.

The zero mode of Virasoro algebra is an important operator
$$
L_0 = \sum_{r\in\Z+\hf} r:\psi_{r}\psi^*_{-r}:
$$
States represented by partitions are eigenstates of $L_0$ with eigenvalues related to the number of boxes of a partition and a charge
\be
L_0 |p,R\rangle = \big(|R| + \frac{p^2}{2}\big)|p,R\rangle.   \label{L0}
\ee


\section{Characters} \label{app-characters}

In this appendix we present explicit formulas for affine characters following the notation used in \cite{kac} and \cite{cft}.

\subsection{$\chi^{\widehat{u}(1)_N}_j$ characters}

$\chi^{\widehat{u}(1)_N}_j$ characters are defined as
\begin{equation}
\chi^{\widehat{u}(1)_N}_j(x) = \frac{1}{\eta(q)}\sum_{n\in\mathbb{Z}} q^{\frac{N}{2}(n+j/N)^2} x^{n+j/N}.  \label{u1-char}
\end{equation}


\subsection{$\chi^{\widehat{su}(k)_1}_r$ characters} \label{subsec-suN-char}

In general, the character of an affine integrable weight $\widehat{\lambda}$ at a given level $l$ and finite part $\lambda$ can be written in terms of string functions $c^{\widehat{\lambda}}_{\widehat{\lambda}'}$ and $\Theta$-functions
\be
\chi^{\widehat{\lambda}}(\zeta,q) = \sum_{\widehat{\lambda}'} c^{\widehat{\lambda}}_{\widehat{\lambda}'} \Theta_{\widehat{\lambda}',\,l}, \qquad \quad  \Theta_{\widehat{\lambda}',\,l}(\zeta,q) = \sum_{\alpha^{\vee} \in Q^{\vee}} e^{2\pi i (l\alpha^{\vee}+\lambda'|\zeta)} q^{\frac{l}{2}|\alpha^{\vee}+\lambda'/l |^2}  \label{aff-char}
\ee
where $\zeta=\sum z_i \alpha_i^{\vee}$ is certain specialization point and $q=e^{2\pi i\tau}$. 

For $\widehat{su}(k)_1$ there is a single string function $c^{\widehat{\lambda}}_{\widehat{\lambda}}=\eta(q)^{-k+1}$, and there are $k$ integrable weights at level 1 with corresponding characters labelled by $r=0,\ldots,k-1$. To be more precise, we consider in fact \emph{specialized} characters, computed at a particular point in Cartan subalgebra. Choosing this point as  $\zeta=\sum z_i \alpha_i^{\vee}$, the characters of $\widehat{su}(k)_1$ take an explicit form
$$
\chi^{\widehat{su}(k)_1}_r(z_i,q) = \frac{\Theta_{\lambda_r}^{level\ 1}(\zeta,q)}{\eta(q)^{k-1}} = \frac{1}{\eta(q)^{k-1}} \sum_{\alpha^{\vee} \in Q^{\vee}} e^{2\pi i (\alpha^{\vee}+\lambda_r|\zeta)} q^{\frac{1}{2}|\alpha^{\vee}+\lambda_r |^2}.
$$
For $r$'th weight we can choose e.g. $\lambda_r = r\omega_1$. For an arbitrary element of coroot lattice $\alpha^{\vee} = \sum_{i=1}^{k-1} n_i \alpha^{\vee}_i$ we get
\begin{eqnarray}
\frac{1}{2}|\alpha^{\vee}+\lambda_r |^2 & = & \sum_i (n_i^2 - n_i n_{i+1})+n_1 r +\frac{r^2}{2}\frac{k-1}{k},  \nonumber \\
e^{2\pi i(\alpha^{\vee}+\lambda_r|\zeta)} & = & y_1^{2n_1-n_2+r} y_2^{2n_2-n_1-n_3}\ldots y_{k-1}^{2n_{k-1}-n_{k-2}} = \prod_{i=1}^{k-1} \tilde{y}_i ^{n_i + \frac{k-i}{k}r}, \nonumber
\end{eqnarray}
where we introduce
$$
y_i=e^{2\pi i z_i},
$$
\begin{equation}
\tilde{y}_1 = \frac{y_1^2}{y_2},\quad \tilde{y}_2 = \frac{y_2^2}{y_1 y_3},\ldots,\quad \tilde{y}_{k-1} = \frac{y_{k-1}^2}{y_{k-2}}.  \label{char-su-vars}
\end{equation}
With such a notation the characters read
\bea
\chi^{\widehat{su}(k)_1}_r(y_i,q) & = & \frac{1}{\eta(q)^{k-1}}\sum_{n_1,\ldots,n_{k-1}} q^{\sum_i (n_i^2 - n_i n_{i+1})+n_1 r +\frac{r^2}{2}\frac{k-1}{k}} y_1^r \prod_{i=1}^{k-1} y_i ^{\sum_j A_{ij}n_j} = \nonumber \\
& = &  \frac{1}{\eta(q)^{k-1}}\sum_{n_1,\ldots,n_{k-1}} q^{\sum_i (n_i^2 - n_i n_{i+1})+n_1 r +\frac{r^2}{2}\frac{k-1}{k}} \prod_{i=1}^{k-1} \tilde{y}_i ^{n_i + \frac{k-i}{k}r}. \label{su-char}
\eea



\subsection{$\chi^{\widehat{u}(k)_1}$ character and its decomposition}

\begin{figure}[htb]
\begin{center}
\includegraphics[width=0.45\textwidth]{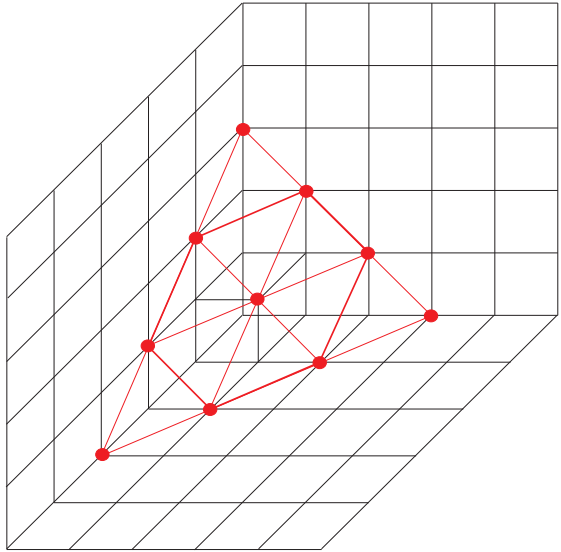}
\begin{quote}
\caption{Decomposition of $\mathbb{Z}^3$ lattice into $\mathbb{Z}$ (along the diagonal $x=y=z$) and $su(3)$ root lattice $Q_{su(3)}$ (in red).} \label{fig-decompose}
\end{quote}
\end{center}
\end{figure}

$\widehat{u}(k)_1$ character is given by a trace over a Fock space of $k$ free complex fermions. It depends on variables $z_i,\ i=1,\ldots,k$ which couple to Cartan currents $J_i$
\begin{eqnarray}
\chi^{\widehat{u}(k)_1}(x_i,q) & = & Tr_{\mathcal{F}}\Big(e^{2\pi i \sum_i z_i J_i}\,q^{L_0 - \frac{k}{24}} \Big) = \nonumber \\
& = & q^{-\frac{k}{24}}\prod_{i=1}^{k} \prod_{p\in\mathbb{Z}_+ + \frac{1}{2}} (1+x_i q^p)(1+x_i^{-1}q^p) = \nonumber \\
& = & \frac{1}{\eta(q)^k}\sum_{\vec{p} = (p_1,\ldots,p_k)\in \mathbb{Z}} q^{\frac{1}{2}(p_1^2 + \ldots + p_k^2)}\, x_1^{p_1}\ldots x_k^{p_k} = \frac{\Theta_{\mathbb{Z}^k}(q;z_i)}{\eta(q)^k}, \label{u-k-1}
\end{eqnarray}
where $x_{i}=e^{2\pi i z_{i}}$. Here the Jacobi triple product identity has been used, and in the last line $\Theta_{\mathbb{Z}^k}$ function is defined in terms of summation over $\mathbb{Z}^k$ lattice. We now show this lattice decomposes as a product of a one-dimensional lattice and $su(k)$ root lattice $Q_{su(k)}$
$$
\mathbb{Z}^k = \sum_{r=0}^{k-1} \mathbb{Z} \times Q_{su(k)},
$$
and rearrange summations appropriately. One-dimensional $\mathbb{Z}$ factor corresponds to $u(1)$ overall charge, so it is spanned by a diagonal $\vec{p}=(n,\ldots,n) \in \mathbb{Z}^k$. For a fixed point on this diagonal , $su(k)$ lattice is given by a perpendicular hyperplane which is spanned by $k-1$ vectors $\epsilon_j-\epsilon_{j+1}$. An example of this decomposition is shown in figure (\ref{fig-decompose}). To probe all points of the original $\mathbb{Z}^k$ in fact an additional shift $r=0,\ldots,k-1$ has to be introduced, and we have to sum over all values of $r$ each point in $\mathbb{Z}^k$ is uniquely specified by a set of numbers $(n;n_1,\ldots,n_{k-1};r)$:
\be
\vec{p}=(p_1,\ldots,p_k) = 
\left[\begin{array}{l}
n+n_1+r \\ 
n-n_1+n_2 \\
n-n_2+n_3 \\
\quad \vdots \\
n-n_{k-2}+n_{k-1} \\
n-n_{k-1}\end{array}\right] \in\mathbb{Z}^k. \label{p-decompose}
\ee

Rewriting the above character in terms of these variables we get
\begin{eqnarray}
\chi^{\widehat{u}(k)_1} & = & \frac{1}{\eta(q)^k}\sum_{r=0}^{k-1} \sum_{n;n_1,\ldots,n_{k-1}} q^{\frac{k n^2}{2} + \sum_i (n_i^2 - n_i n_{i+1}) +\frac{r^2}{2} + nr + n_1 r} x_1^{n+n_1+r}x_2^{n+n_2-n_1}\ldots x_k^{n-n_{k-1}} = \nonumber \\
& = & \frac{1}{\eta(q)^k}\sum_{r=0}^{k-1} \sum_{n} q^{\frac{k}{2}(n+r/k)^2} \tilde{x}^{n+r/k} \sum_{n_1,\ldots,n_{k-1}} q^{\sum_i (n_i^2 - n_i n_{i+1})+n_1 r +\frac{r^2}{2}\frac{k-1}{k}} \prod_{i=1}^{k-1} \tilde{x}_i ^{n_i + \frac{k-i}{k}r} \label{uNchar}
\end{eqnarray}
where we introduced new variables
\begin{equation}
\tilde{x} = x_1 x_2\cdots x_k;\qquad \tilde{x}_i = \frac{x_i}{x_{i+1}},\ i=1,\ldots,k-1. \label{char-u-vars}
\end{equation}
Comparing with (\ref{u1-char}) and (\ref{su-char}), this can be written as 
\begin{equation}
\chi^{\widehat{u}(k)_1}(x_i) = \sum_{r=0}^{k-1} \chi^{\widehat{u}(1)_k}_r(\tilde{x})\, \chi^{\widehat{su}(k)_1}_r(\tilde{x}_i) \label{char-u-su}
\end{equation}
where $r$ runs over different $u(1)$ charges and $\widehat{su}(k)_1$ weights.


\end{document}